\documentclass[aps,prl,twocolumn,superscriptaddress]{revtex4}

\usepackage{natbib}
\usepackage{graphicx}
\usepackage{subfigure}
\usepackage{setspace}
\usepackage{umoline}
\usepackage{color}

\definecolor{rltred}{rgb}{0.75,0,0}
\definecolor{rltgreen}{rgb}{0,0.5,0}
\definecolor{oneblue}{rgb}{0,0,0.75}
\definecolor{marron}{rgb}{0.64,0.16,0.16}
\definecolor{forestgreen}{rgb}{0.13,0.54,0.13}
\definecolor{purple}{rgb}{0.62,0.12,0.94}
\definecolor{dockerblue}{rgb}{0.11,0.56,0.98}
\definecolor{freeblue}{rgb}{0.25,0.41,0.88}
\definecolor{myblue}{rgb}{0,0.2,0.4}

\begin{document}

\title{Fast path and polarisation manipulation of telecom wavelength single 
photons in lithium niobate waveguide devices}
\author{Damien Bonneau}
\affiliation{Centre for Quantum Photonics, H. H. Wills Physics Laboratory \& Department of Electrical and Electronic Engineering, University of Bristol, Merchant Venturers Building, Woodland Road, Bristol, BS8 1UB, UK}
\author{Mirko Lobino}
\affiliation{Centre for Quantum Photonics, H. H. Wills Physics Laboratory \& Department of Electrical and Electronic Engineering, University of Bristol, Merchant Venturers Building, Woodland Road, Bristol, BS8 1UB, UK}
\author{Pisu Jiang}
\affiliation{Centre for Quantum Photonics, H. H. Wills Physics Laboratory \& Department of Electrical and Electronic Engineering, University of Bristol, Merchant Venturers Building, Woodland Road, Bristol, BS8 1UB, UK}
\author{Chandra M. Natarajan}
\affiliation{Scottish Universities Physics Alliance and School of Engineering and Physical Sciences, Heriot-Watt University, Edinburgh, EH14 4AS, United Kingdom}
\author{Michael G. Tanner}
\affiliation{Scottish Universities Physics Alliance and School of Engineering and Physical Sciences, Heriot-Watt University, Edinburgh, EH14 4AS, United Kingdom}
\author{Robert H. Hadfield}
\affiliation{Scottish Universities Physics Alliance and School of Engineering and Physical Sciences, Heriot-Watt University, Edinburgh, EH14 4AS, United Kingdom}
\author{Sanders N. Dorenbos}
\affiliation{Kavli Institute of Nanoscience, TU Delft, 2628CJ Delft, The Netherlands}
\author{Val Zwiller}
\affiliation{Kavli Institute of Nanoscience, TU Delft, 2628CJ Delft, The Netherlands}
\author{Mark G. Thompson}
\affiliation{Centre for Quantum Photonics, H. H. Wills Physics Laboratory \& Department of Electrical and Electronic Engineering, University of Bristol, Merchant Venturers Building, Woodland Road, Bristol, BS8 1UB, UK}
\author{Jeremy L. O'Brien}
\email{Jeremy.OBrien@bristol.ac.uk}
\affiliation{Centre for Quantum Photonics, H. H. Wills Physics Laboratory \& Department of Electrical and Electronic Engineering, University of Bristol, Merchant Venturers Building, Woodland Road, Bristol, BS8 1UB, UK}
\date{\today}

\begin{abstract}
We demonstrate fast polarisation and path control of photons at 1550~nm in lithium niobate waveguide devices using the electro-optic effect. We show heralded single photon state engineering, quantum interference, fast state preparation of two entangled photons and feedback control of quantum interference. These results point the way to a single platform that will enable the integration of nonlinear single photon sources and fast reconfigurable circuits for future photonic quantum information science and technology.
\end{abstract}

\maketitle

Photons are an appealing approach to quantum information science, which promises enhanced information and communication technologies \cite{ob-nphot-3-687}. An integrated optics approach appears essential for practical applications as well as advances in the fundamental science of quantum optics \cite{po-jstqe-15-1673,njp-special-issue}. Progress towards single photon sources \cite{sps,sh-nphot-1-215}, detectors \cite{ha-nphot-3-696}, and circuits that use path \cite{po-sci-320-646,po-sci-325-1221,pe-sci-329-1500,pe-ncomm-2-224,la-apl-97-211109} 
and polarisation \cite{sa-prl-105-200503} encoding have been made, including circuits that are reconfigurable to allow manipulation of photon paths \cite{ma-nphot-3-346,sm-oe-17-13516}. However, these reconfigurable circuits have relied on inherently slow thermal phase shifters and operation at 800~nm. Fast operation of reconfigurable waveguide circuits at telecom wavelengths is crucial for 
integration with the existing optical telecom networks as well as to benefit from the technologies developed in that area.

Fast routing and manipulation of single photons is essential for both temporally and spatially multiplexed single photons sources \cite{mi-pra-66-053805,ma-pra-83-043814,je-jmo-58-276,mc-prl-103-163602}, quantum communication, including device independent quantum key distribution \cite{gi-rmp-74-145}, based on noiseless linear amplifiers \cite{xi-nphot-4-316}, circuit \cite{kn-nat-409-46} and measurement \cite{ni-prl-93-040503,br-prl-95-010501,pr-nat-445-65} based quantum computing \cite{ob-sci-318-1567}, quantum control \cite{gi-prl-104-080503}, and interaction free measurements \cite{kw-prl-74-4763}. Fast control of both path and polarisation is crucial for these applications. Proof of principle demonstrations of fast manipulation of single photons have been made using bulk Pockels cells \cite{pr-nat-445-65,gi-prl-104-080503,ma-pra-83-043814,kw-prl-74-4763,mc-prl-103-163602}, {however, there have been no such demonstrations in integrated quantum photonic circuits.}

A fast electro-optic effect and the ability to make low loss single mode waveguides at telecom wavelengths using either proton exchange or titanium indiffusion (Ti:LN) makes lithium niobate an appealing platform for fast reconfigurable quantum photonic devices. Lithium niobate is used in telecommunications applications where 40 GHz modulators are standard; 100 GHz has been demonstrated in the laboratory \cite{Atsushi_Kanno2010817}. Polarization controllers based on the electro-optic effect have also been demonstrated for bright light \cite{Haasteren1993,ngw-jlt-8-438}. Lithium niobate is also appealing for the prospect of directly integrating periodically poled waveguide photon sources \cite{am-njp-12-103005}. 

\begin{figure*}
   \includegraphics[width=170mm]{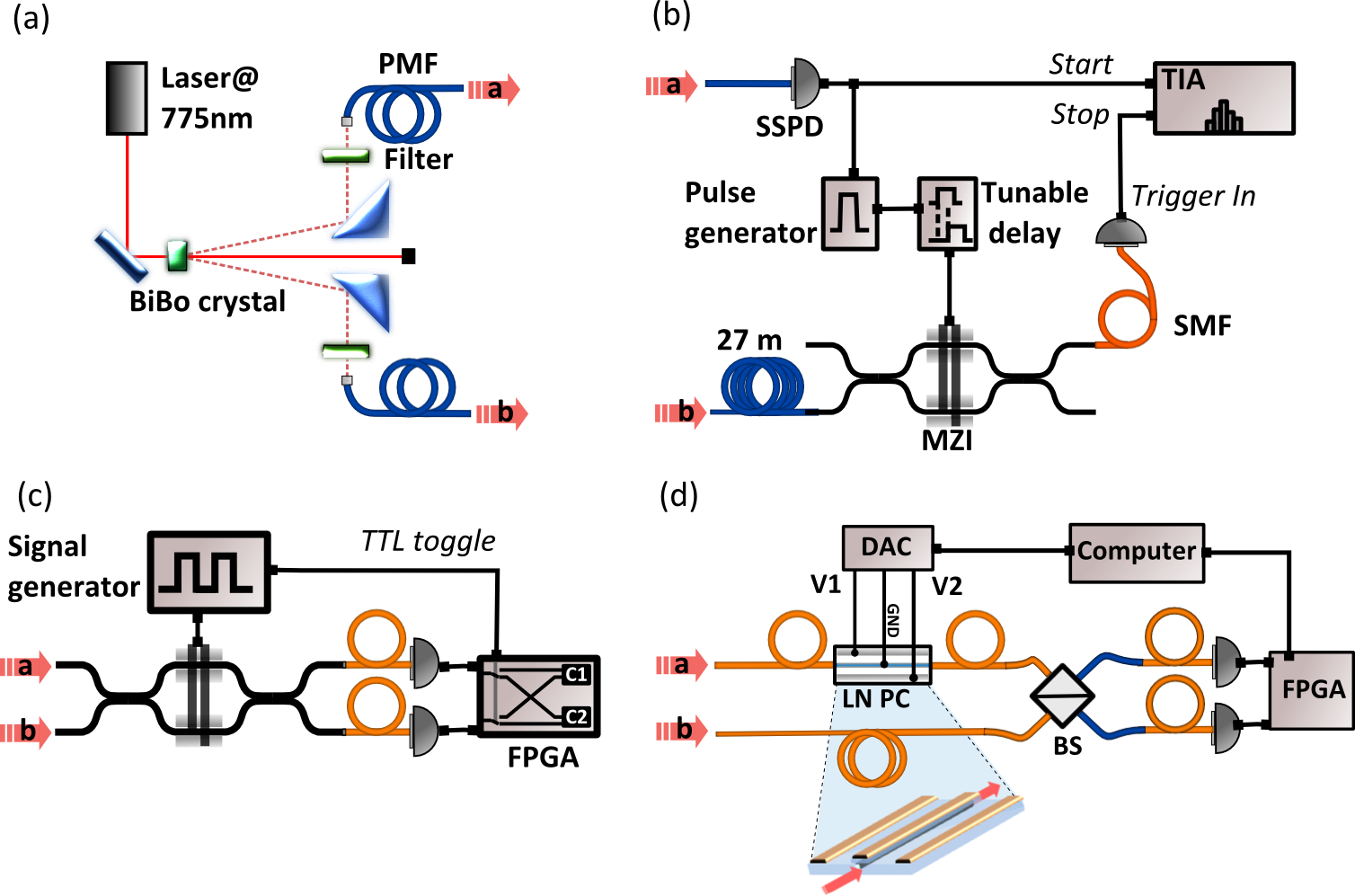}
    \caption{
    \label{MZI}
   {Fast path and polarisation control of single photons in lithium niobate waveguides.}
{(a) Pulsed spontaneous parametric downconversion (SPDC) source for photon pairs at 1550~nm. (b) Fast heralded single photon state preparation set-up. MZI: lithium niobate Mach-Zehnder interferometer; SSPD: single photon superconducting detectors; SMF: single mode fibre; TIA: time interval analyzer. (c) Fast switching of two photon entangled state. The signal generator (SG) drives the MZI with square waves alternating between two voltages $V_0=-1.6V$ and $V_\frac{\pi}{2}=0.5V$. Coincidental events for the two voltages are recorded in separate counters embedded in a field programmable gate array (FPGA) board. (d) Fast polarisation feedback control of single photons using a lithium niobate polarisation controller (PC). The inset shows one stage of the PC consisting of a waveguide surrounded by 3 electrodes. 
}}
\end{figure*}


Fast manipulation of photon path is possible using the device shown in Fig.~\ref{MZI}(b): A Mach-Zehnder Interferometer (MZI) fabricated in Ti:LN is composed of two 50:50 directional couplers. Electrodes above each waveguide inside the MZI enable rapid manipulation of the refractive index via the eletro-optic effect. Application of the same positive voltage $V$ to each of the inside electrodes relative to the outside ground electrodes produces an equal and opposite electric field, and hence change in refractive index and phase, in each arm of the MZI. 

The same Ti:LN waveguide technology can be used to control the polarisation of a single photon when it is integrated with an appropriate electrode architecture \cite{Haasteren1993,ngw-jlt-8-438}. The inset to Fig.~\ref{MZI}(d) shows a schematic of {one stage of} the electro-optical polarization controller (PC) {used in} this work: application of voltages $V_1$ and $V_2$ on the electrodes either side of the waveguide, relative to the ground electrode above the waveguide enables an arbitrary electric field vector to be applied perpendicular to the waveguide. The device therefore acts as a tunable waveplate with a controllable thickness $\rho$ and rotation $\varphi$ that realises the rotation ${\hat{R}}=exp[{\frac{2\pi}{\lambda}\rho\left(\mathbf{\hat{\sigma}_x}\sin 2\varphi+\mathbf{\hat{\sigma}_z}\cos 2\varphi  \right)}]$. {The PC has 4 identical stages (one is shown in the inset of Fig.~\ref{MZI}(d)) which enables the implementation of any unitary operation of the single photon polarisation when they are controlled independently.}

Photon pairs at 1550~nm wavelength were generated by spontaneous parametric down conversion (SPDC) in a bismuth borate (BiBO) crystal and collected into two polarization maintaining optical fibres (see Fig.~\ref{MZI}(a)), analogous to the $\sim$800 nm SPDC sources that have been used routinely over the last decades. Single photons were detected with two superconducting single photon detectors (SSPDs) \cite{ha-oe-13-10846,na-apl-96-211101,do-apl-93-131101,ta-96-221109} having system detection efficiencies of 8\% and 18\%, respectively (see Appendix for further details).

\begin{figure}
   \includegraphics[width=85mm]{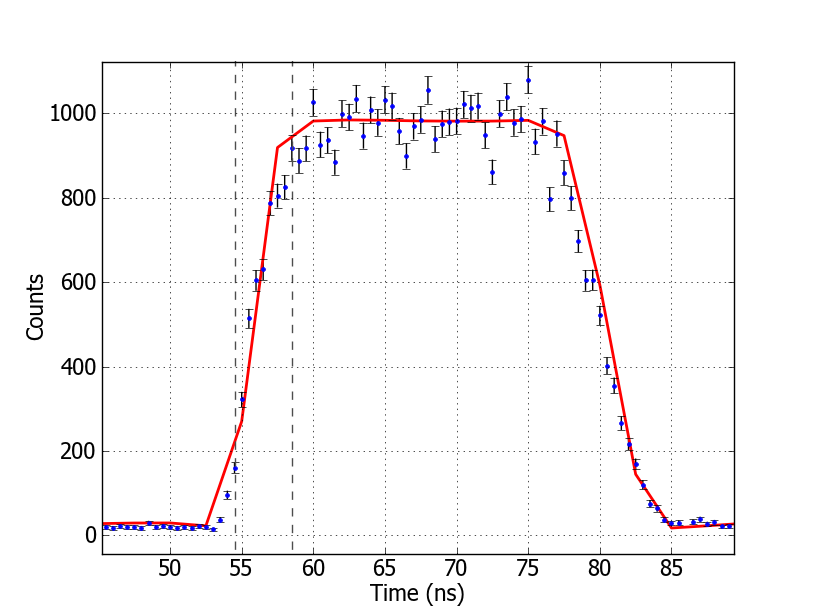}
    \caption{
    \label{singlephotonswitchplot}
   {Fast path control of single telecom wavelength photons in a lithium niobate Mach-Zehnder interferometer.}
    Number of coincidence events (accumulated in 30~s) as a function of the delay between the optical and the electric pulse (Fig. 1(b)).
The red line is the expected switching behaviour computed from the shape of the pulse and the classical characterisation of the MZI with a DC voltage. The dots are the measured values. The error bar associated are $\pm\sigma$ for Poissonian statistics.
    }
\end{figure}

Figure~\ref{MZI}b shows the experimental set-up used for heralded single photon state preparation. One photon is measured directly by an SSPD providing the trigger signal for the pulse generator that controlled the MZI. For every trigger event, a voltage pulse was sent to the MZI with a controllable delay. This pulse induced a relative phase shift $\theta$ which performed the transformation $\left|10\right\rangle\rightarrow\sin\mbox{\ensuremath{\frac{\theta}{2}}\ensuremath{\ensuremath{\left|10\right\rangle }-\ensuremath{\cos\frac{\theta}{2}}}}\left|01\right\rangle$.
We drove the MZI with a voltage pulse of 20~ns duration and 4~ns rise time that switched between $\theta=\pi$, corresponding to the identity transformation, to $\theta=0$, for the swap transformation which routed the photon to the second SSPD. By measuring the number of heralded counts as a function of the delay applied to the driving pulse we reconstructed the time response of the interferometer. Figure~\ref{singlephotonswitchplot} shows the number of heralded single photon events as a function of the pulse arrival time, incremented in 0.5~ns steps. The switching efficiency is 97.9$\pm$0.1\% with a switching time of 4~ns limited by the waveform of the driving voltage.

\begin{figure}
   \includegraphics[width=\columnwidth]{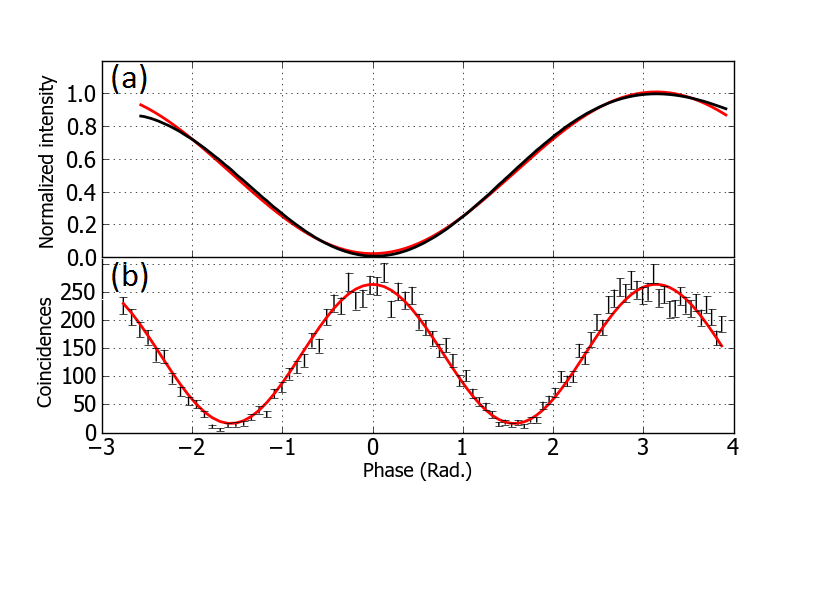}
    \vspace{-2cm}
    \caption{
    \label{twoPhotonSwitch}
   {Phase control of a two photon state.}
    (a) Classical interference fringe showing the intensity at one output of the MZI as a function of applied voltage.
    (b) Two-photon interference fringe showing the number of photon pairs detected simultaneously at each output of the MZI in 40~s.
Each dot represents experimental data and the red line is a squared sinusoidal fit.
The error bars are $\pm \sigma$ for Poissonian statistics.
    }
\end{figure}

A continuous range of two-photon states can also be prepared with the reconfigurable MZI: Injecting a pair of photons into its two input ports and applying a voltage $V_\theta$ implements the transformation $\left|11\right\rangle\rightarrow\frac{\sin\theta}{\sqrt{2}}\mbox{\ensuremath{\left(\mbox{\ensuremath{\left|20\right\rangle }}-\mbox{\ensuremath{\left|02\right\rangle }}\right)}}-\cos\theta\mbox{\ensuremath{\left|11\right\rangle }}$. We verified the tunability of the MZI by continuously changing the voltage from -5~V to +5~V and measuring the two photon coincident events at the two output ports. From this measurement we retrieved the $\left|11\right\rangle$ component of the prepared states and the associated 2-photon fringe shown in Fig.~\ref{twoPhotonSwitch}(b), together with the bright light (single photon) fringe (Fig.~\ref{twoPhotonSwitch}(a)). The two-photon fringe has a visibility $V_{2ph}=95.2\pm1.4$\% and half the period of the single photon fringe. 
The non-unit visibility is attributed primarily to imperfect spectral overlap of the photons.

\begin{figure}
   \includegraphics[width=80mm]{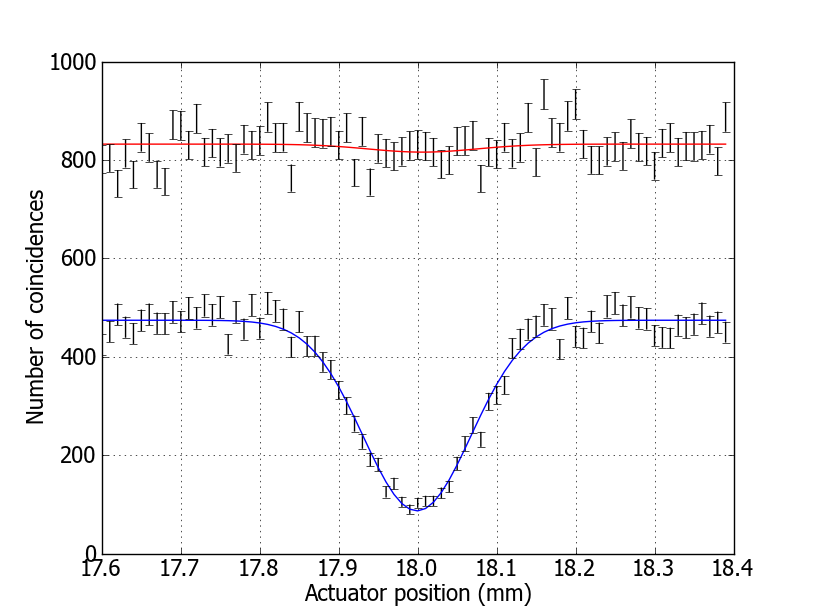}
    \caption{
    \label{twophotonfastswitchdata}
   {Fast switching of a two photon state.}	
	By delaying the arrival time of one photon with respect to the other (shown on the x axis), we simultaneously record the coincidental counts integrated over 210~s (y axis) for each voltage applied to the MZI. In the case where the applied voltage is close to $V_\frac{\pi}{2}=0.5V$, the MZI acts as a balanced beamsplitter, quantum interference occurs and a dip of $82\pm2$\% visibility is recorded. For $V_0=-1.6V$, the MZI acts as a crosser, therefore, the photons do not interfere and a visibility of $2\pm3$\% is recorded.}
\end{figure}

\begin{figure*}
   \includegraphics[width=170mm]{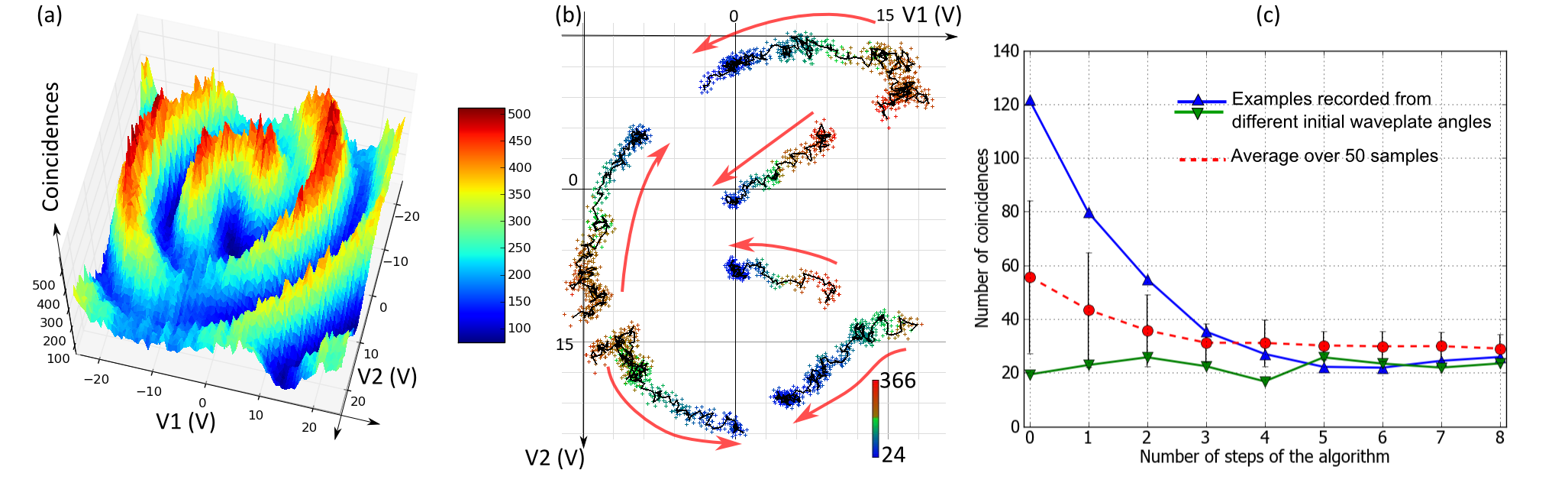}
    \caption{
    \label{scan3Dpolcontrol}
  {Polarisation control of 1550~nm photons, using the device and setup shown in Fig. 1c.}
    {(a) Coincident photon counts in 4~s as a function of applied voltages $V_1$ and $V_2$. (b) Paths generated by applying the feedback loop algorithm starting from a random $V_1$ and $V_2$. Colored  points represent the measurement of the number of coincidences integrated over 4~s while black lines show the path followed by the algorithm. The arrows indicate the direction of the evolution for the six different initial conditions. (c) Dynamic of the feedback loop, showing the number of coincidences as a function of the number of steps of the algorithm. The blue and green lines are two examples recorded from different initial waveplate angles. The dashed line is the average of 50 random samples.}
    }
\end{figure*}

Fast two-photon state preparation was realized by applying a 4~MHz square wave that alternates between the voltages $V_0$ and $V_{\pi/2}$ (see Fig.~\ref{MZI}(c)). With this driving signal the interferometer continuously switches between the output states $\left|11\right\rangle$ and $\left(\mbox{\ensuremath{\left|20\right\rangle }}-\mbox{\ensuremath{\left|02\right\rangle }}\right)/\sqrt{2}$. Two separate counters C1 and C2, embedded on the same electronic board, were used to record coincidental events arising from the $V_0$ and $V_{\pi/2}$ settings, respectively. 
During this measurement we varied the relative delay between the two photons by translating one of the two collection fibres of the SPDC source with a motorized stage. As we changed this delay, we recorded two simultaneous count rates for $V_0$ and $V_{\pi/2}$ as shown in Fig.~\ref{twophotonfastswitchdata}. For $V_0$ the state
$\left|11\right\rangle$ is ideally prepared and no dependence of the coincidence count rate on the delay was observed. In contrast, for $V_{\pi/2}$
when we measured the number of coincidences as a function of the relative delay between the two photons 
we observed the expected Hong-Ou-Mandel interference dip \cite{HOM87} with a visibility $V=82\pm2$\%. 

The driving electronics of the MZI are understood to be a major contributor to this non-unit visibility since the data shown in Fig.~\ref{twoPhotonSwitch}(b) with fringe visibility $V_{2ph}=95.2\pm1.4$\% corresponds to a Hong-Ou-Mandel interference dip visibility of $V=91.2\pm$2.6\%. The capacitance of the modulator induces a pseudo-periodic voltage oscillation that causes phase oscillations, and could be reduced by optimising the driving electronics.


{To control the polarisation of single photons we connected the four stages of the PC (Fig.~\ref{MZI}(d))} in parallel such that only two driving voltages $V_1$ and $V_2$ were required. We used this reconfigurable ``integrated waveplate" in an active feedback loop implemented to maximize the polarisation indistinguishability between two single photons interfering at a 50/50 beamsplitter (BS) (Fig.~\ref{MZI}(d)). Assuming that the state of the photons arriving at the BS is $\left|\psi\right\rangle_{1} \otimes\left|\phi\right\rangle_{2} =\left(\cos\alpha\left|H\right\rangle_{1} +e^{i\beta}\sin\alpha\left|V\right\rangle_{1} \right)\otimes\left(\cos\gamma\left|H\right\rangle_{2} +e^{i\delta}\sin\gamma\left|V\right\rangle_{2} \right)$ and that the polarization drift of the photons is slow compared to the time required to complete a full feedback loop it is possible to align their polarisation by implementing the transformation $\hat{R}$ satisfying $\hat{R}\left|\psi\right\rangle_{1}= \left|\phi\right\rangle_{2}$ with our integrated polarization controller (PC).

Figure~\ref{MZI}(d) shows the experimental set-up: here the photons from the SPDC source were collected into two single mode optical fibres that were not polarisation maintaining.
One photon was sent through the polarisation controller and the other through a fiber patchcord to arrive simultaneously at the beamsplitter. We first characterized the PC by fixing the two collection fibres to the table and measuring the number of coincidences from the two outputs of the BS as a function of $V_1$ and $V_2$. Figure~\ref{scan3Dpolcontrol}(a) shows the measured coincidence pattern with a visibility of the quantum interference $V_{pol}=87\pm1\%$. Maximum coincidental event detection was observed when the two photon polarisations were orthogonal while low coincidental detection corresponded to identical polarisation; non-unit visibility is mainly due to multi-photon events that arise from a higher pump power of 400 mW (95\% visibility was observed in a conventional Hong-Ou-Mandel experiment at the 10 mW power used for all other 2-photon demonstrations reported here).

Next we used the polarisation controller to automatically optimize the quantum interference between two single photons which were nominally identical in all degrees of freedom except for polarization. For this task there is no need to perform tomography of any of the states since any minimum in the number of coincidences is the global minimum (see Appendix). Because of this property of the coincidence function we implemented a feedback loop based on the gradient descent method. Figure~\ref{scan3Dpolcontrol}(b) shows the convergence of six coincidence paths towards the minimum,  starting from six different random polarizations. In all cases the system evolves towards the minimum coincidences condition which implies maximum indistinguishability between the photons.

The dynamic response of the feedback loop was measured by periodically changing the polarization of one photon via the rotation of a computer controlled (bulk) half waveplate placed before the collection fibre of the SPDC source. This setup simulates the situation where a single photon propagating in a controlled environment interferes with a second photon coming from a noisy channel. In this situation the PC is used to compensate for polarization fluctuation of the second photon and restore maximum indistinguishability.
We quantified the level of indistinguishability restored by the feedback loop by measuring the average number of coincidences. 
The overlap between the polarisations of the two photons is restored after $\sim$4 iterations (Fig.~\ref{scan3Dpolcontrol}(c)). The speed of the loop was limited by the low coincidence count rate of the SPDC system that required an integration time of 2~s in order to acquire a meaningful number of events.

Rapid manipulation of the polarisation and path degrees of freedom of single photons will be essential for future quantum technologies as well as fundamental quantum science. The ability to perform both path and polarisation manipulation in a single platform is particularly appealing. Furthermore, lithium niobate promises the ability to directly integrate periodically poled LN single photon sources. Ultimately it should also be possible to integrate SSPDs into the waveguide circuit via growth of NbTiN directly onto LN substrates \cite{do-apl-93-131101}. A particularly important future application is multiplexed single photon sources\cite{mi-pra-66-053805,ma-pra-83-043814,je-jmo-58-276,mc-prl-103-163602}: The setup shown in Fig. 1(a,b) represents a single unit of such a source: by removing the pump beam block and replicating the BiBO crystal and MZI $N$ times, simple switching logic would enable a near-deterministic single photons source to be realised, provided efficiencies and losses could be controlled; a fully integrated architecture will help reduce such losses. (We note that heralding efficiency reported here is not state-of-the art.) Reconfigurable circuits with path and polarisation encoding will find applications across photonic quantum information science and technology ranging from quantum communication, quantum control, quantum measurement and quantum information processing.
\\
\\We thank O. Alibart and S. Tanzilli. This work was supported by Nokia, EPSRC, ERC, PHORBITECH, QUANTIP, NSQI, FOM and NWO (Vidi grant). M.L. acknowledges a Marie Curie IIF. R.H.H. acknowledges a Royal Society University Research Fellowship. J.L.O'B. acknowledges a Royal Society Wolfson Merit Award.





\bibliography{bib15b,biblitest}

\clearpage

\section*{Appendix}
\subsection*{1550~nm photon source}
A BiBO crystal ($\Theta = 8.8^\circ, \Phi = 0^\circ$, 4~mm thickness), was pumped by a Ti-Sapphire pulsed laser at 775~nm wavelength, focused on the crystal by a plano-convex lens (f = 30~mm). The pulse width was around 80~fs and the repetition rate was 80~MHz. Photon pairs at 1550~nm wavelength arising from spontaneous parametric down conversion were filtered by 10~nm width bandpass filters and collected by aspheric lenses (f = 11~mm). The source was tested at 30~mW power and exhibited 95\% visibility quantum interference when the two arms were combined on a balanced beam splitter.

\subsection*{Quantum interference from a two photon state with arbitrary polarisations}

We start with an unlimited number of copies of two unknown state of polarisation $\left|\Psi_{1}\right\rangle$  and $\left|\Psi_{2}\right\rangle$ , and assume that we have control of the parameters of $\left|\Psi_{1}\right\rangle$. Using an iterative process, we aim to achieve quantum interference between the two states without having to measure the polarisation of any of the states.

The two photon input state can be written as $\left|\Psi_{In}\right\rangle =\left|\Psi_{1}\right\rangle \otimes\left|\Psi_{2}\right\rangle =\left(c_{\alpha}\left|H\right\rangle +e^{i\beta}s_{\alpha}\left|V\right\rangle \right)\otimes\left(c_{\gamma}\left|H\right\rangle +e^{i\delta}s_{\gamma}\left|V\right\rangle \right)$

where $c_{x}$ and $s_{x}$ are respectively $\cos(x)$ and $\sin(x)$

Writing the creation operators as $\mathbf{m_{X}^{\dagger}}$ where m is the spatial mode and X the polarisation, the input state can be rewritten as :

$\left|\Psi_{In}\right\rangle =\left(c_{\alpha}\mathbf{a_{H}^{\dagger}}+e^{i\beta}s_{\alpha}\mathbf{a_{V}^{\dagger}}\right)\otimes\left(c_{\gamma}\mathbf{b_{H}^{\dagger}}+e^{i\delta}s_{\gamma}\mathbf{b_{V}^{\dagger}}\right)\mathbf{\left|\mathbf{0}\right\rangle }$

The beam splitter scattering matrix provides :
$
\begin{array}{ccc}
\mathbf{a_{X}^{\dagger}}&\rightarrow&\frac{\mathbf{c_{X}^{\dagger}}+i\mathbf{d_{X}^{\dagger}}}{\sqrt{2}}\\
\mathbf{b_{X}^{\dagger}}&\rightarrow&\frac{i\mathbf{c_{X}^{\dagger}}+\mathbf{d_{X}^{\dagger}}}{\sqrt{2}}
\end{array}
$
Combining the two photons $\left|\Psi_{1}\right\rangle$ and $\left|\Psi_{2}\right\rangle$  on a beam splitter gives the output state

$
\begin{array}{ccc}
\left|\Psi_{out}\right\rangle &=&\frac{1}{2}\left[c_{\alpha}c_{\gamma}\left(\mathbf{c_{H}^{\mathbf{2}\dagger}}-\mathbf{\mathbf{d}_{H}^{2\dagger}}\right)+e^{i(\beta+\delta)}s_{\alpha}s_{\gamma}\left(\mathbf{c_{\mathbf{V}}^{\mathbf{2}\dagger}}-\mathbf{\mathbf{d}_{\mathbf{V}}^{2\dagger}}\right)\right]
\\&+&\frac{i}{2}\left(c_{\alpha}s_{\gamma}e^{i\delta}+c_{\gamma}e^{i\beta}s_{\alpha}\right)\left(\mathbf{\mathbf{c_{H}^{\dagger}}c_{V}^{\dagger}}+\mathbf{\mathbf{d}_{H}^{\dagger}}\mathbf{\mathbf{d}_{V}^{\dagger}}\right)
\\&+&\frac{1}{2}\left(c_{\alpha}s_{\gamma}e^{i\delta}-c_{\gamma}e^{i\beta}s_{\alpha}\right)\left(\mathbf{\mathbf{c_{H}^{\dagger}}d_{V}^{\dagger}}-\mathbf{\mathbf{c}_{V}^{\dagger}}\mathbf{\mathbf{d}_{H}^{\dagger}}\right)
\end{array}
$
And the probability of a 2-photon coincidental detection is then given by

$P_{coinc}=\frac{1}{2}\left|c_{\alpha}s_{\gamma}e^{i\delta}-c_{\gamma}e^{i\beta}s_{\alpha}\right|^{2}$

We are interested in using an algorithm to minimize the number of coincidences. And for this purpose, we need to find where the gradient is 0. The derivative of $P_{coinc}$ with respect to the parameters $\alpha$ and $\beta$ on which we have control are given by :

$
\begin{array}{ccc}
\frac{\partial P_{coinc}}{\partial\alpha}&=&\frac{1}{2}\left[\sin\left(2\alpha\right)\cos\left(2\gamma\right)-\cos\left(2\alpha\right)\sin\left(2\gamma\right)\cos\left(\delta-\beta\right)\right]\\\frac{\partial P_{coinc}}{\partial\beta}&=&\frac{1}{2}\left[\sin\left(\beta-\delta\right)\sin\left(2\alpha\right)\sin\left(2\gamma\right)\right]
\end{array}
$

The gradient is 0 if and only if $\frac{\partial P_{coinc}}{\partial\alpha}=0$ and $\frac{\partial P_{coinc}}{\partial\beta}=0$. This condition is satisfied only in the following cases:


$
\begin{array}{ccc}
\left|\Psi_{1}\right\rangle =\left|H\right\rangle$ or $\left|V\right\rangle &$and$&\left|\Psi_{2}\right\rangle =\left|H\right\rangle$ or $\left|V\right\rangle
\\
&$or$&
\\
\beta=\delta+n\pi&$and$&\alpha=\gamma+k\frac{\pi}{2}
\end{array}
$

where $\left\{k,n\right\} \in Z^{}$  

The first case provides either identical or orthogonal photons. The second case gives $\left|\Psi_{1}\right\rangle =\left|\Psi_{2}\right\rangle$  up to a global phase for even k and $\left|\Psi_{1}\right\rangle \perp\left|\Psi_{2}\right\rangle$  for odd k. The only cases where the gradient is 0 are only either when the two photons are the same up to a global phase or when they are orthogonal. Thus, in all cases, any local minimum is also the global minimum and any local maximum is also the global maximum.

\subsection*{Superconducting single-photon detector system}

In this study we used a pair of high performance nanowire superconducting single-photon detectors (SSPDs) \cite{go-apl-79-705} integrated into a practical closed cycle detector system \cite{ha-oe-13-10846}.  We have previously validated the use of such detectors in quantum waveguide circuit experiments at 805~nm \cite{na-apl-96-211101}.   The SSPDs used in this study were NbTiN nanowire meanders fabricated on oxidised silicon substrates \cite{do-apl-93-131101}, coupled efficiently with single mode optical fibre.  The packaged SSPDs were operated at 3~K in a Gifford-McMahon type closed cycle refrigerator \cite{ha-oe-13-10846}.  The system detection efficiency of our two SSPD channels was 18\% and 8\% at 1550~nm wavelength and an ungated dark count rate per detector of 1~kHz.  The measured timing jitter of each detector channel was 60~ps full width at half maximum.

\end{document}